\documentclass{article}
\usepackage{amsmath}
\usepackage{float}
\usepackage{fullpage}
\usepackage{graphicx}
\usepackage{hyperref}
\usepackage{apalike}

\usepackage[round]{natbib}
\usepackage{url}
\usepackage{hyperref}\hypersetup{colorlinks=false,urlcolor=blue}
\usepackage[document]{ragged2e}
\usepackage{parskip}
\usepackage{endnotes}

\makeatletter
\renewcommand{\maketitle}{\bgroup\setlength{\parindent}{0pt}
\begin{flushleft}
\textbf{\@title}\\\bigskip
  \@author
  \end{flushleft}\egroup
}
\makeatother

\usepackage{multicol,color}

\begin{document}

\title{\Large Network Analysis with the Enron Email Corpus}

\author{\hyperref[jo]{\color{blue}\underline{J.~S.\ Hardin}} \\ \hyperref[gs]{\color{blue}\underline{G.\ Sarkis}} \\ \hyperref[pc]{\color{blue}\underline{P.~C.\ URC}}\\Pomona College}


\maketitle

\section*{Abstract} We use the Enron email corpus to study relationships in a network by applying six different measures of centrality.  Our results came out of an in-semester undergraduate research seminar.  The Enron corpus is well suited to statistical analyses at all levels of undergraduate education. Through this article's focus on centrality, students can explore the dependence of statistical models on initial assumptions and the interplay between centrality measures and hierarchical ranking, and they can use completed studies as springboards for future research.  The Enron corpus also presents opportunities for research into many other areas of analysis, including social networks, clustering, and natural language processing.

\section{Introduction}

One of the most infamous corporate scandals of the past few decades curiously left in its wake one of the most valuable publicly-available datasets.  In late 2001, the Enron Corporation's accounting obfuscation and fraud led to the bankruptcy of the large energy company.  The Federal Energy Regulatory Commission subpoenaed all of Enron's email records as part of the ensuing investigation.  Over the following two years, the commission released, unreleased, and rereleased the email corpus to the public after deleting emails that contained personal information like social security numbers.  The Enron corpus contains emails whose subjects range from weekend vacation planning to political strategy talking points, and it remains the only large example of real world email datasets available for research.  See \citet{FERC} for the Federal Energy Regulatory Commission's website on the Enron investigation, \citet{DataRelease} for the final order releasing the data to the public, and \citet{Smarties} for a popular account of the Enron scandal.

Research into the corpus is prolific and wide ranging.  We present here a selection from the large range of publications on Enron to highlight some of the research that the corpus has spurred, and to suggest possible further directions as well.
See \citet{EReport} for a technical report describing a MySQL database of the corpus, \citet{TimeSeries} for anomaly detection in a dynamic network, \citet{Diesner05} for a social network analysis that focused on changes in behavior during the scandal period, \citet{Deitrick12} for a neural networks model predicting the gender of an emailer based on the email stream,  \citet{Peterson2011} for measures of formality in the email correspondence,  \citet{Spectral} for a graph-theoretic and spectral analysis that overlaps with many of the topics of interest in our article, \citet{Martin} for detection of abnormal email activity in outgoing messages, \citet{Zhou06} for a probabilistic approach to community detection, and \citet{ZhouAlias} for data cleaning with focus on email aliases.

\subsection{Network Analysis}
The network of communication between Enron employees naturally induces a graph whose nodes are labeled by employees and whose edges correspond to email communication.  We weight  the edge between the two nodes by the number of emails sent.  Additionally, we use directionality to separately analyze emails sent or emails received, when appropriate.

Networks are ubiquitous in the internet age, underlying much of virtual (and real) life from social webs to recommender systems, and from epidemiological spread to linguistic evolution.  They are used widely as tools of research in sociology \citep{Sut14} (patterns of Tweets during a natural disaster), biology \citep{Pin11} (coordinated behavior of harvester ants), genetics \citep{Zha05} (co-expressed gene groups in brain cancer), and economics \citep{Ste10} (economic value of a social network in a large online marketplace) to study the behavior of individuals and of systems.

We discuss six measures of the Enron corpus based on the adjacency matrices of the email network, and we suggest how they can be used in undergraduate education and research.  We also provide a brief analysis of the group membership of the most connected cliques, found by hierarchical clustering.  Our results and methods came out of an undergraduate research circle at Pomona College that we oversaw during the spring semester of 2014.  The research circle consisted of four students whose interests and initiative determined the research questions and research direction, and two math/stats faculty members who provided general and technical guidance.

See Section \ref{subsection:data} for details on how the matrices were constructed, Section \ref{sec:res} for the research questions and some of the results, and Section \ref{sec:centrality} for a survey of the centrality measures. In Sections \ref{subsecHH} and \ref{fd}, we suggest ways that our research project can be incorporated into the undergraduate curriculum.

\section{Dataset Story: Cleanup and Processing}\label{subsection:data}
The narrative aspect of many datasets in both pedagogy and research includes a major data-collection component.  Even in classroom examples where the data, or a summary thereof, is given to the students, there often exists a contextual story about how and why the data might have been collected for the immediate purpose of the statistical analysis.  The Enron corpus, on the other hand, is for all intents and purposes an accidental, incidental dataset.  This presents an invaluable opportunity to discuss real-world data issues that do not often come up in the classroom.  Specifically, real-world data is often dirtier and less cooperative than experimental data.  It is not structured with a specific goal in mind---it is what it is. Therefore, getting it to the tidy stage where analysis may be conducted and meaning may be extracted involves several assumptive and simplifying decisions that require thoughtful analysis before the fact (see, for example, Hadley Wickham's work on the vital aspect of tidying data \citep{Wickham14}).  Additionally, the Enron dataset is clearly observational and provides much fodder for a classroom discussion on the limits of inferences done on observational data.

\begin{figure}[H]
\begin{center}
\includegraphics[scale=.6]{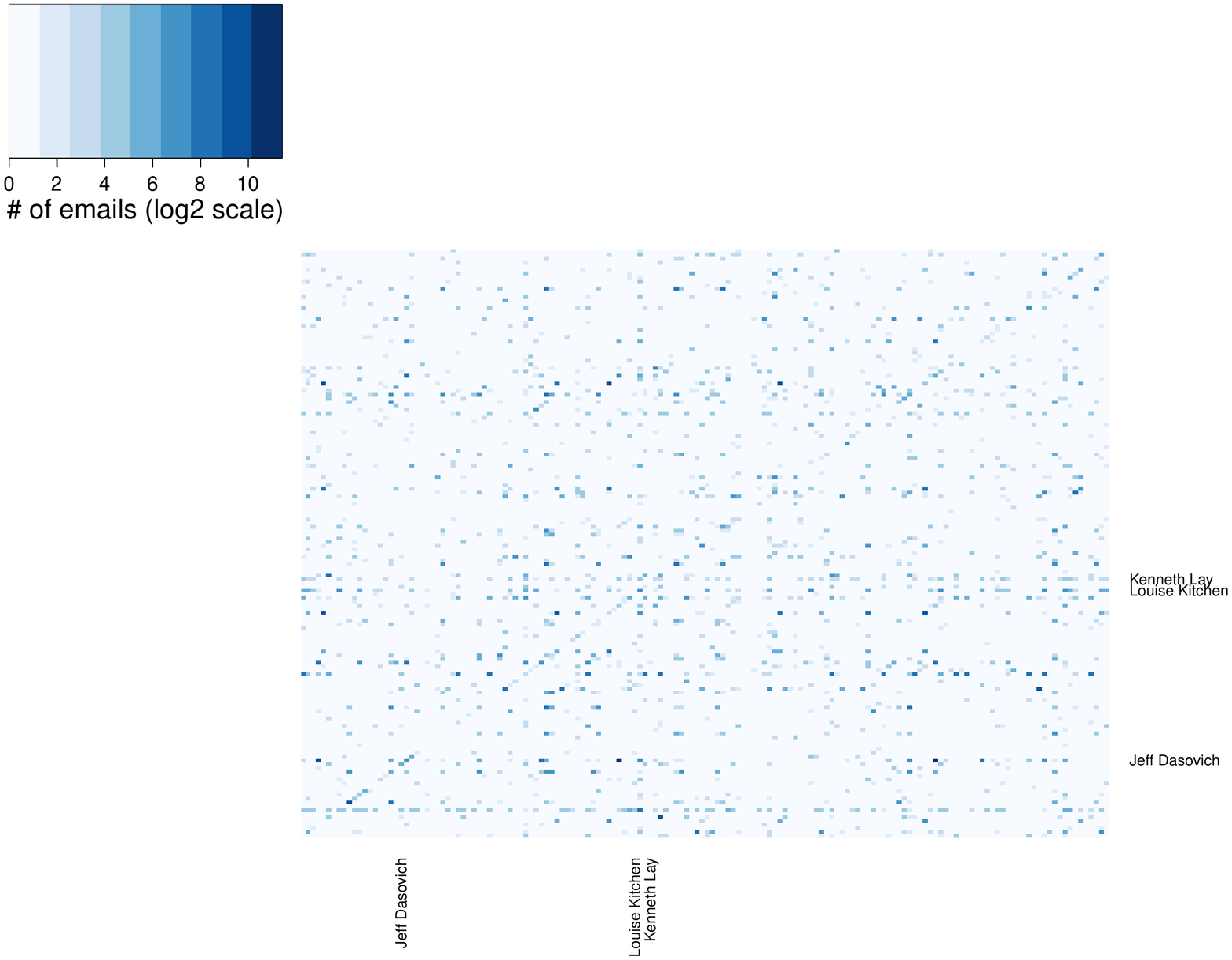}\\
\caption{\label{dirmat} Each $(i,j)^{th}$ dot represents a binary indicator that an email was sent from person $i$ to person $j$.  White indicates no communication, while the darker the color, the more communication between person $i$ and person $j$.  Note that for the rows, the $i^{th}$ individual counts up from the bottom.  E.g., Jeff Dasovich is the $21^{st}$ column and the $21^{st}$ row {\em counting from the bottom}}
\end{center}
\end{figure}

For our project, we used the dataset available at \url{https://s3.amazonaws.com/metanautix/enron/enron_mail_20110402_csv.tgz}, whose emails were organized into 150 mailboxes labeled by employee name; the emails in a mailbox were not necessarily sent by that person. Additionally, some employees with similar names were binned into the same mailbox, while others had their messages split among two mailboxes.  In order to circumvent such potential binning errors, we ignored the folder designation and instead extracted only {\sl From}, {\sl To}, and {\sl CC} \ fields of each email message.  While only one employee may appear in either the {\sl From} or {\sl To} fields (which is different from most current email systems), an arbitrary number may appear in the {\sl CC} field. We considered only senders and recipients with email addresses that have an enron.com domain name.  To distinguish between the individuals, we relied on six standard aliases used at Enron (see \citet{ZhouAlias} for instance). The result was 156 employees whose email communication we considered, and from which we constructed an adjacency matrix for the weighted directed graph of Enron employees, as visualized in figure \ref{dirmat}.  The dots in the figure are colored so that the darker the color of the point at the ($i,j$) entry, the more emails were sent from person $i$ to person $j$.

Let $E_{ij}$ be the set of emails for which Enron employee $i$ appears in the {\sl From} field and employee $j$ appears in the {\sl To} field.  Let $C_{ij}$ be the set of emails for which Enron employee $i$ appears in the {\sl From} field and employee $j$ appears in the {\sl CC} field.  For each $c\in C_{ij}$, let $n_c$ be the number of names that appear in the {\sl CC} field of $c$.  Define the $156\times 156$ weighted adjacency $M$ as: \begin{equation}m_{ij}=|E_{ij}|+\sum_{c\in C_{ij}}\frac{1}{\sqrt{1+n_c}} \label{data} \end{equation}
Thus, for the weighting of the edge in the directed graph from employee $i$ to employee $j$, each email sent from $i$ to $j$ contributed $1$, and each email $c$ sent from $i$ on which $j$ was cc-ed contributed $1/\sqrt{1+n_c}$.

We considered the contribution of cc-ed emails to be less important than emails sent directly; in asking ``how many cc-ed emails is one direct email worth?'', we arrived at a square-root relationship. Our answer came out of a discussion with the students; different groups may reach different conclusions regarding the appropriate weighting.  This is a valuable opportunity for the students to explicitly consider the consequences of the assumptions they make.

The matrix $M$ is the weighted adjacency matrix of a directed graph.  To consider the undirected graph, we defined the matrix $U=M+M^T-D$, where $D$ is the diagonal matrix $d_{ii}=2m_{ii}$.  In other words, for the undirected graph, we did not incorporate information from emails that employees cc-ed themselves on.  In subsection \ref{subsecAA}, Alternative Applications, we discuss choices within the data cleanup process for creating the adjacency matrix we used as well as alternative adjacency matrices.

While the network of 156 nodes is relatively small in size, its edges included more than 500,000 email messages and 18GB of data; by way of reference, we note that a movie can range from 1GB in standard definition to 6GB in Blu-ray.  This made the cleanup component of our investigation a big data project.  The methods of aggregating the data are outside the scope of this article; we imported the data into MySQL and used simple queries to count how many emails employee $i$ sent to employee $j$, and iterated the queries over all pairs of employees.  However, it is worth noting that the construction of the $156\times 156$ matrix took several days of computation, and that different considerations in computing the entries (as discussed below) can provide interesting alternate research routes into the data.

Our students also used D3, a java-based library, to visualize the network.  Their work is at \\
\url{http://obscure-meadow-3612.herokuapp.com/} and at \url{http://enron-network.herokuapp.com/TOM}.

\section{Research questions}\label{sec:res} We are interested in the social network that is defined by the emails.  In particular, we investigate what kinds of information about the relative importance of the Enron employees can be read from the graph whose vertices are the employees and whose edges represent email correspondence.  To that end, we consider six measures of centrality based on the email network: degree, eigencentrality for sent emails, eigencentrality for received emails, closeness, betweenness, and topological overlay, which we discuss in more detail in Section \ref{sec:centrality} below.

There is no reason to believe that the kinds of importance rankings induced by an email connectivity graph reflect the managerial structure of the corporation itself. Indeed, rankings based on email networks point to an overlay of the activity of an individual emailer and the subnetwork of contacts that emailer has. As such, centrality measures based on an email network may help gauge the {\sl functional} importance of various employees, as opposed to (or in conjunction with)  their managerial importance; and different centrality measures are more adept at spotting different functionalities, which we give examples of next.

Though the scope of our project was primarily exploratory, we did observe some interesting results that may form the basis for further research directions.  Consider Table \ref{tab:ranks} below, which summarizes the ranks of the top 10 employees for each of the size centrality measures.
\begin{itemize}
\item The overlap between the six top-ten lists is not insignificant: 29 employees make up the 60 names.  Two employees appear in five of the six lists: Jeff Dasovich, the director for state government affairs, ranks highly in each but the eigencentrality measure based on received emails; and Louise Kitchen, an energy trader in the European market and COO of Enron Wholesale Services, ranks highly in all but topological overlap.
\item On the other hand, 14 of the employees appear just once.  Among these are Kenneth Lay, the Chairman of Enron, who ranked seventh in betweenness, and Greg Whalley, its president, who ranked eighth in closeness.  The only other board member of Enron to appear on any of the lists is Stephen Kean, Vice President and Chief of Staff, who ranked sixth in degree and third in topological overlap.
\item Counting multiplicity, 25 of the 60 employees who ranked in the top ten were legal counsels of some kind, either in the Enron North America Legal Department (21) or otherwise having ``Counsel" in their job title (4).  In other words, the email network captured the importance of the legal departments at Enron.
\item Thirteen of the 29 unique employees who ranked in the top ten were women, and, counting multiplicity, 33 of the 60 were women; in comparison, 38 of the total 156 employees were women.
\end{itemize}

\begin{table}[ht]
\centering
\begin{tabular}{rllllll}
  \hline
 & Degree & EVcent & EVcentT & Closeness & Betweenness & TOM \\
  \hline
1 & Jeff Dasovich & Tana Jones & Sara Shackleton & Robert Benson & Louise Kitchen & Jeff Dasovich \\
  2 & Mike Grigsby & Sara Shackleton & Susan Bailey & Mike Grigsby & Mike Grigsby & Richard Shapiro \\
  3 & Tana Jones & Stephanie Panus & Marie Heard & Louise Kitchen & Susan Scott & Steven J. Kean \\
  4 & Sara Shackleton & Marie Heard & Tana Jones & Kevin M. Presto & Jeff Dasovich & Mike Grigsby \\
  5 & Richard Shapiro & Susan Bailey & Stephanie Panus & Susan Scott & Mary Hain & Tana Jones \\
  6 & Steven J. Kean & Kay Mann & Elizabeth Sager & Scott Neal & Sally Beck & Sara Shackleton \\
  7 & Louise Kitchen & Louise Kitchen & Jason Williams & Barry Tycholiz & Kenneth Lay & Mary Hain \\
  8 & Susan Scott & Elizabeth Sager & Louise Kitchen & Greg Whalley & Scott Neal & Marie Heard \\
  9 & Michelle Lokay & Jason Williams & Jeffrey T. Hodge & Phillip K. Allen & Kate Symes & Stephanie Panus \\
  10 & Chris Germany & Jeff Dasovich & Gerald Nemec & Jeff Dasovich & Cara Semperger & Susan Scott \\
   \hline
\end{tabular}
\caption{According to each of six different measures of centrality, we provide a ranked list of the individuals who are most central to the email corpus.}\label{tab:ranks}
\end{table}

\section{Centrality and Rank}\label{sec:centrality}

A measure of centrality on a graph aims to assign a ranking or magnitude to each node that captures the relative importance of that node in the context of the graph's structure.    We are interested in measuring the importance of each employee based on the number of emails sent or received, as aggregated in the dataset we extracted from the Enron corpus and summarized in the matrices $M$ and $U$.  Recall that $M$ is the weighted $156\times 156$ adjacency matrix of our directed graph,  and $U$ is the corresponding weighted matrix of the undirected graph that does not distinguish between sent and received emails.

We investigate six measures of importance within the Enron employee email network: degree, eigenvector centrality for received emails, eigenvector centrality for sent emails, closeness, betweenness, and topological overlay.  We give an overview of the measures below, including mathematical definitions and intuition.  For each of the measures, it may be of interest for students to generate examples of nodes in a network that rank high or low in centrality.  In Section \ref{fd} below, we also suggest how some of the measures may be incorporated into statistics classes of various levels.

We make two general observations about the employees who ranked highly according to the centrality measures.  First, while `Vice President,' `Director,' and `President' appear frequently in their titles, only three of these employees were Enron board members.  The centrality rankings therefore captured a functional participation in the email network rather than managerial importance.  Second, while there was nontrivial overlap between and correlation among the lists, each of the centrality measures seemed to pick out a distinctive narrative feature from among the employees.

\subsection{Degree \iffalse and strength\fi} The {\sl degree} $\delta_i$ of employee $i$ is defined to equal the total number of employees to whom $i$ sent or received emails.  Thus, if we define $$\delta_{ij}=\begin{cases} 1&\text{ if } u_{ij}\neq 0\\0&\text{ if } u_{ij}=0\end{cases}$$then $\delta_i=\sum_j\delta_{ij}$.  We did not distinguish between whether $j$ appeared in the {\sl To} or {\sl CC} field.  The degree is a measure of the size of $i$'s immediate network.  The more different people $i$ emails, directly or by cc, or receives emails from, the greater $i$'s degree.

The top-ranked employee according to degree centrality is Jeff Dasovich, the Director of Regulatory and Government Affairs.  Note that the only top-ten list Jeff Dasovich does not appear in is the transpose-eigencentrality one, suggesting that his presence in the other top-ten lists is on the strength of his emails sent rather than received.  Also, the Enron departments are well represented in this list---there are seven unique departments among the ten employees.  That is, there does not appear to be one department clearly more active than the others in email communications based on count alone.

Note that $\delta$ is the adjacency matrix of the unweighted, undirected graph.
It may be of interest to compute the degree using the weighted and/or directed matrix instead, so that the results might be comparable to other measures below using the matrices $M$ and $U$.  

\subsection{Eigenvector Centrality} Denote the centrality of employee $i$ with the nonnegative real number $x_i$. Suppose $x_i$  is accumulated from the centralities $x_j$ as $j$ ranges over all employees that $i$ emails.  Suppose further that employee $j$ contributes to $x_i$ in direct proportion to the connectedness from $i$ to $j$ as measured by $m_{ij}$.  That is,
$$x_i=\frac{1}{\lambda}\sum_j m_{ij}x_j\text{\quad where }\frac{1}{\lambda}\text{ is some proportionality constant}.$$
While the definition appears circular (the centrality $x_i$ depends on $x_j$, which in turn depends on $x_i$), we can summarize the relationships with the familiar matrix equation $M\vec{x}=\lambda\vec{x}$, where $\vec{x}=\begin{pmatrix}x_1&\cdots&x_{156}\end{pmatrix}^T$.  In other words, $\vec{x}$ is an eigenvector of $M$ with eigenvalue $\lambda$.  While $M$ may have several different eigenvalues and eigenvectors, the Perron-Frobenius Theorem (see \citet[page 508]{MA90} for instance) guarantees that because $m_{ij}\geq 0$, then for some eigenvalue $\lambda$ with largest absolute value, there exists an eigenvector $\vec{x}$ whose entries are all nonnegative; that so-called dominant eigenvector provides the importance weights for the employees.

We also consider $M^T$, the transpose of $M$, in order to analyze importance based on received emails.  In that case, we compute $M^T\vec{x}=\lambda\vec{x}$ for the eigenvector of importance weights corresponding to the eigenvalue $\lambda$ with the largest absolute value.

Such a measure of importance is called {\sl eigenvector centrality}.  While it is commonly used with binary or stochastic matrices, its premise applies to any matrix with nonnegative entries.  Arguably the most famous instance of eigenvector centrality is the first implementation of Google's PageRank algorithm: webpages rank highly in Google's search results if they are linked from other webpages of high rank.  See \citet{PageRank} for a fun illustration of the algorithm suitable for a linear algebra class, and \citet{LarrySergey} for the original paper by Google founders Sergey Brin and Lawrence Page.

An employee ranks highly in eigenvector centrality (respectively, transpose-eigenvector centrality) if that employee sends emails to (respectively, receives emails from) many other highly-ranked employees.  Students can generate and discuss examples of company structure based on whether the two eigenvector centralities turn out to be highly correlated or uncorrelated: what kinds of employees might rank highly in eigenvector centrality of emails sent, but not in the transpose of emails received?

Both top-ten lists pertaining to eigenvector centrality have a high legal representation---7 for emails sent and 8 for emails received.  So even though the legal department appears only twice in the degree top-ten list, its emails must have been sent from/to more central people as measured by the eigenvector equation.

Moreover, 8 of the 10 in eigencentrality-sent and 7 of the 10 in eigencentrality-received are women.  In both lists combined, there is only one legal department employee who is not a woman (Jeffrey T. Hodge, who appears in only one list), and there is only one woman who is not an employee of the legal department (Louise Kitchen, who is tied with Jeff Dasovich for appearing in the most top-ten lists, five of the six).  For comparison, women make up around half of the legal department and a quarter of the total 156 employees considered.

\subsection{Closeness} Given a pair of employees $i$ and $j$, a {\sl path} from $i$ to $j$ is defined to be a sequence $i_0=i,i_1,\cdots,i_r=j$ such that $m_{i_{t-1}i_t}\neq 0$ for all $1\leq t\leq r$. A path from $i$ to $j$ is called a {\sl shortest path} if it minimizes the number of steps $r$ in the sequence, and the {\sl distance} from $i$ to $j$, denoted $d(i,j)$, is the number of steps in such a shortest path from $i$ to $j$.  The {\sl closeness} of $i$ is then defined as $$\gamma_i=\frac{1}{\sum_{j\neq i} d(i,j)}$$
If the graph has more than one connected component---in other words, if there exists a pair of nodes that cannot be connected via a sequence of edges---then the closeness of any node equals zero.  Otherwise, the closeness of $i$ measures the speed or efficiency with which information spreads out from $i$ to the rest of the graph.  Note that $\gamma_i$ is sometimes normalized by the number of nodes other than $i$ so that it measures the reciprocal of the average distance from $i$: $(n-1)/\sum_{j\neq i} d(i,j)$. However, the ranking of employees based on closeness is independent of such a normalization.

An employee has high closeness centrality if that employee's correspondence reaches a large proportion of the network quickly.  Thus closeness is a measure of the entire network's structure in relation to a node.  It may be interesting to discuss the robustness of the closeness centrality.  For instance, can an employee rise in the rankings by sending one or two carefully chosen emails?

According to the Enron graph, six of the top ten employees with respect to closeness centrality were directors or vice presidents of trading.  The remaining four are Susan Scott, one of two women and the only lawyer in the group, Jeff Dasovich and Louise Kitchen, each of whom appeared in five of the six top-ten lists, as well as Greg Whalley, Enron's President.

\subsection{Betweenness}  For betweenness, we consider the undirected network and its adjacency matrix $U$, and ask path-related questions similar to those in closeness measures.  Given a pair of employees $j$ and $k$, an {\sl undirected path} from $j$ to $k$ is defined to be a sequence $i_0=j,i_1,\cdots,i_r=k$ such that $u_{i_{t-1}i_t}\neq 0$ for all $1\leq t\leq r$; in particular, we do not take into account the from/to directedness of the graph of Enron employees. A path from $j$ to $k$ is called a {\sl shortest undirected path} if it minimizes the number of steps $r$ in the sequence.

Since our paths are weighted, we make an adjustment \citep{Newman, Opsahl} to the definition of ``shortest'' under the premise that a larger weight implies a closer connection between the two corresponding employees, and in that sense, a shorter path.  Let $i_0,i_1,\cdots,i_r$ be a path from $i_0$ to $i_r$.  Then the weighted length of the path is the reciprocal sum of the weights of the path's edges: $$\sum_{1\leq t\leq r}\frac{1}{u_{i_{t-1}i_t}}.$$

Let $\tau_{jk}$ be the number of shortest undirected paths from $j$ to $k$, and let $\tau_{jk}(i)$ be the number of paths from among those in $\tau_{jk}$ that pass through $i$.  Then the {\sl betweenness} of $i$ is given by $$\beta_i=\sum_{i\neq j\neq k}\frac{\tau_{jk}(i)}{\tau_{jk}}$$
As such, the betweenness of $i$ measures the importance of $i$ as a central node in efficient communication between other nodes in the network. An employee has high betweenness centrality if that employee figures prominently in the email proximity of many pairs of colleagues.

The directed/undirected choices for closeness/betweenness, respectively, naturally generate discussion questions about the reasons for the choices and about how the measures might differ if alternate choices were made.  As well, students can investigate different modifications for the shortest path in closeness to account for weighting.

Perhaps the most noteworthy aspect of the betweenness top ten is a single appearance, and by association, a single absence. Kenneth Lay, one of the two characters most commonly associated with the Enron scandal, comes in at number 7. This is the only appearance of Kenneth Lay in any of the top ten lists.  Also, Jeffrey Skilling, the other face of the scandal, does not appear on any of the top-ten lists.

\subsection{Topological Overlap Matrix}

Topological Overlap Matrix (TOM) extends the adjacency matrix from a measure of connectedness between two nodes only to a measure of connectedness between two nodes and the rest of the individuals in the dataset \citep{Rav02,Yip07}.  Let $u_{ij}$ be the measure of adjacency between nodes $i$ and $j$ as defined in Subsection \ref{subsection:data}.  We define the matrix $TOM$ as:
$$ TOM_{ij} = \frac{\sum_{l \ne i,j} u_{il} u_{lj} + u_{ij}}{\min\bigg(\sum_{l \ne i,j} u_{il}, \sum_{l \ne i,j} u_{jl}\bigg) +1 - u_{ij}}$$
This new adjacency matrix is then converted to a centrality measure by taking the row sum of the $TOM$.  That is, the most central node will be the one who is most connected to the other nodes by way of third party connections.  It is worth pointing out that TOM directly accounts for the second degree connections, and so it will naturally produce different measures of importance than other centrality measures.

Topological overlap adjacency was originally designed to take as input unweighted networks, or binary matrices.  In such a case, $TOM_{ij}$ measures the proportion of overlap between $i$'s and $j$'s immediate neighbors.  There are three natural avenues of TOM-discussion for students.  First, one may ask  if using the same formula for weighted networks, as we did, can result in misleadingly inflated entries in the TOM matrix---for instance, should the quadratic growth of  $\sum_{l \ne i,j} u_{il} u_{lj}$ be tempered with a square root?  Second, students can consider the Generalized Topological Overlap Matrix \citep{Yip07} which measures the the network overlap of all neighbors within a fixed distance from any two nodes.  And third, the row-sum measure of TOM centrality is a direct measure of the second order connections from the degree centrality; as such, the other measures of centrality discussed above can be used within the TOM matrix to evaluate their second order connections as well.

The TOM top-ten list shares 7 employees in common with the degree list.  This correlation can be seen for all the 156 employees in Figure \ref{fig:pairs}.  The list includes 5 lawyers and 3 executive members of the Regulatory and Government Affairs department.  The remaining two are Mike Grigsby, a vice president of trading, and Steven Kean, Enron's chief of staff, and one of only three board members to be ranked by our centrality measures.

\subsection{Clustering and Network Cliques}
Instead of ranking employees individually, we could ask whether certain groups of employees acted in concert together more so than others.  To do so, we employ hierarchical network construction, as follows.

First, compare all pairs of nodes, or employees, and connect the two nodes which are most similar.  Second, connect the next two nodes which are most similar {\em or} connect a node to the already connected group using either average connectedness, minimum connectedness, or maximum connectedness between the node and that group.  The construction happens iteratively by making one additional connection at each step until all nodes are connected into one group \citep{Eve11}.

\begin{figure}[H]
\begin{center}
\includegraphics[scale=.75]{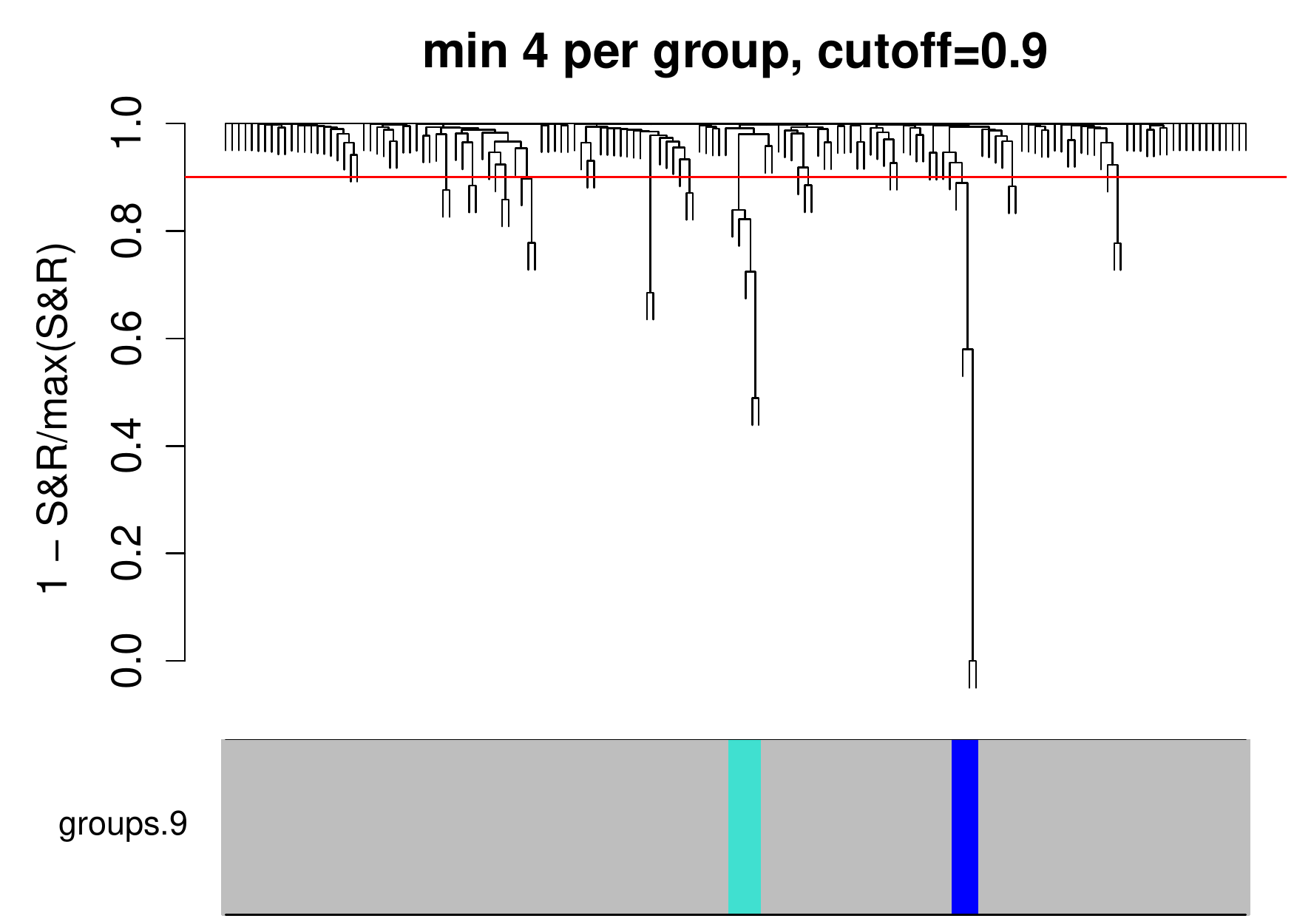}\\
\includegraphics[scale=.75]{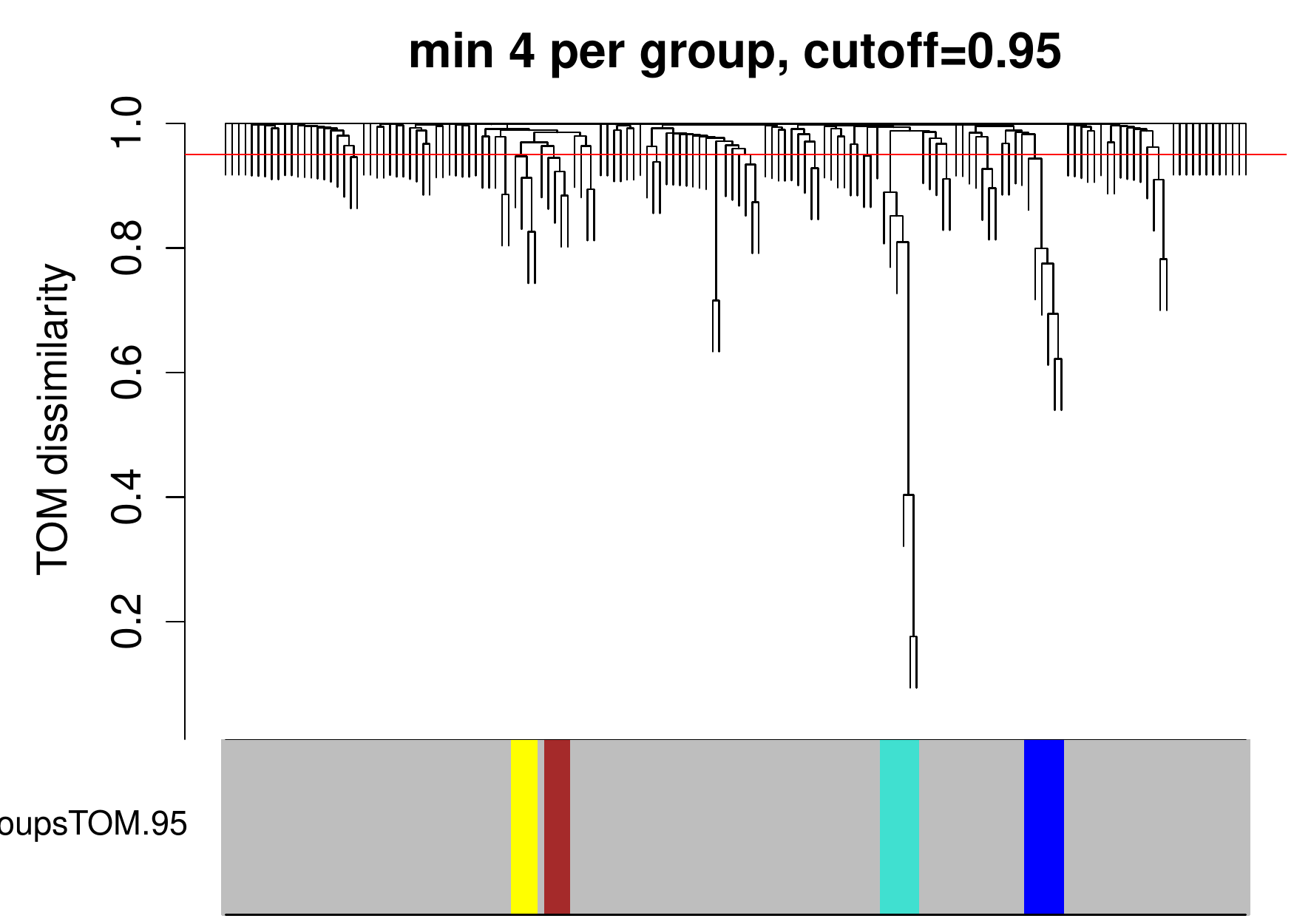}\\
\caption{\label{hierfig} Dendrograms representing hierarchical clustering with the symmetric adjacency matrix (R\&S refers to ``number of emails received and sent") as well as the TOM construction based on the symmetric adjacency matrix.  We group points based on similarity in an agglomerative (bottom up) manner.  Individuals who are similar according to a cutoff (0.9 for symmetric adjacency and 0.95 for TOM) and have at least a minimum cluster size (here 4 individuals) are considered to make up a group.}
\end{center}
\end{figure}

Such bottom-up grouping is called agglomerative, though the splitting mechanism could have happened top-down and would be called divisive.  The result of the splitting algorithm is visualized in a dendrogram (see Figure \ref{hierfig}).  The y-axis of the dendrogram is given by the dissimilarity between any two nodes (or groups of nodes).

Clustering requires a choice for similarity between network nodes.  We used two distances for our clustering. One was based on the number of emails sent and received: the similarity between two nodes equals the proportion of emails sent and received between the respective pair of employees (that is, degree scaled by dividing through by the maximum number of emails sent and received).  The other was based on the TOM matrix: the distance between two groups equals the average TOM distance between all pairs of points across the two groups.

 In our graphs, we define a cluster to be a group of individuals who is both somewhat similar (sent many emails to each other) and has a minimum membership (we arbitrarily set the minimum to be four).   However, hierarchical networks have the disadvantage that in building the network, once two nodes are connected, they remain connected.

Using the number of emails sent and received to measure similarity, we produce two clusters:

\begin{tabular}{lll}
\hline
Name & Department/Title & Centrality (ranking)\\\hline
Susan Bailey & ENA Legal/Legal Specialist & EV (5), EVT (2)\\
Marie Heard & ENA Legal/Legal Specialist & EV (4), EVT (3), TOM (8)\\
Tana Jones & ENA Legal/Legal Specialist & Deg (3), EV (1), EVT (4), TOM (5)\\
Stephanie Panus &  ENA Legal/Legal Specialist & EV (3), EVT (5), TOM (9)\\
Sara Shackleton & ENA Legal/General Counsel Assistant & Deg (4), EV (2), EVT (1), TOM (6)\\\hline\hline
Jeff Dasovich & Reg. and Gov. Affairs/Director & Deg (1), EV (10), Cl (10), Bet (4), TOM (1)\\
Mary Hain  & Reg. and Gov. Affairs/Director & Bet (5), TOM (7)\\
Steven J. Kean & Enron/VP \& Chief of Staff & Deg (6), TOM (3)\\
Richard Shapiro & Reg. and Gov. Affairs/VP & Deg (5), TOM (2)\\\hline
\end{tabular}

Observe that the clusters are remarkably uniform in job title and department.  Additionally, every member of the two clusters apeared in at least two top-ten lists; and the top-ten lists had a lot of overlap (every member of the first cluster ranked in both eigencentralities, and every member of the second cluster ranked highly in TOM).   However, ranking high in centrality measures is not a guarantee of cluster membership; for instance, Louise Kitchen, who appeared in five of the 6 top-ten lists, is not in any of the clusters.

Using the TOM adjacency matrix, we produce four clusters:

\begin{tabular}{lll}
\hline
Name & Department/Title & Centrality (ranking)\\\hline
Susan Bailey & ENA Legal/Legal Specialist & EV (5), EVT (2)\\
Marie Heard & ENA Legal/Legal Specialist & EV (4), EVT (3), TOM (8)\\
Tana Jones & ENA Legal/Legal Specialist & Deg (3), EV (1), EVT (4), TOM (5)\\
Stephanie Panus &  ENA Legal/Legal Specialist & EV (3), EVT (5), TOM (9)\\
Elizabeth Sager & ENA Legal/VP \& General Assistant Counsel & EV (8), EVT (6)\\
Sara Shackleton & ENA Legal/General Counsel Assistant & Deg (4), EV (2), EVT (1), TOM (6)\\\hline\hline
Robert Badeer& ENA West Power/Mgr Trading & none\\
Jeff Dasovich& Reg. and Gov. Affairs/Director & Deg (1), EV (10), Cl (10), Bet (4), TOM (1)\\
Mary Hain& Reg. and Gov. Affairs/Director & Bet (5), TOM (7)\\
Steven J. Kean& Enron/VP \& Chief of Staff & Deg (6), TOM (3)\\
Richard Shapiro& Reg. and Gov. Affairs/VP & Deg (5), TOM (2)\\
James D. Steffes& Reg. and Gov. Affairs/VP &none\\\hline\hline
Lindy Donoho & ETS/Employee & none\\
Michelle Lokay & ETS/Director & Deg (8)\\
Mark McConnell & ETS/Director & none\\
Kimberly Watson & ETS/Director & none\\\hline\hline
Drew Fossum & ETS/VP \& Gen. Cnsl. & none\\
Steven Harris & ETS/VP & none\\
Kevin Hyatt & ETS/Director & none\\
Susan Scott & ETS/Counsel & Deg (8), Cl (5), Bet (4), TOM (1)\\\hline
\end{tabular}
Again observe that the clusters are department-uniform.  The first two TOM clusters include as subsets, respectively, the first two clusters based on number of emails sent/received.  The other two TOM clusters are made up entirely of Enron Technical Services employees, and indeed, of the 12 managerial-level employees in the ETS department, 7 appear in the two last clusters.

\section{Helpful Hints \label{subsecHH}}

Our results build on \citet{Kay14}, a semester long research experience for a group of undergraduates at Pomona College.  We consider the topics to be upper level undergraduate techniques which could easily be taught in a multivariate statistics course, a machine learning computer science course, or a data science course.    Additionally, network analysis or clustering could easily be added as a topic to a course on statistical applications.

We also see a place for the topics in math courses that look for applications to their methods.  In an analysis course that covers metric spaces, networks provide an interesting field of play.  In a linear algebra course, eigenvector centrality can make the mathematical theory come alive.

The use of recent and meaningful data improves the classroom outcomes in terms of both engaging students and solidifying their technical knowledge. It has been our experience that students engage more thoughtfully with statistical methodology when they are interested in the research question at hand---an interest that is usually concurrent with providing intriguing data.  Our experience is in line with the ASA's recently endorsed guidelines promoting exactly this type of meaningful data integration within the undergraduate curriculum \citep{asa14}.  Indeed, in our research circle, the students were given free range to choose both the data set to work with and the analysis method to apply for our semester long research project.  They unanimously chose to work with the Enron corpus and apply network analysis to the email counts.

The Enron corpus is in many ways an ideal dataset for statistical pedagogy.  Although it is not well-suited for standard Neyman-Pearson hypothesis testing, the questions which can be addressed speak to more modern statistical challenges.  There are myriad reasons for using the Enron corpus in a classroom setting: the corpus's origins are unusual and engaging for students who are interested in real-world data and recent American econo-cultural history; a sizeable literature already exists on the corpus, so that students need not start the conversation and investigation at square one; social networks are accessible, especially in the post-Facebook era, yet they motivate current and active research problems; centrality measures are intuitive and mathematically nontrivial; and the discussion presented below may be used for stand-alone research modules in an undergraduate statistics course or may serve as a starting point for a more intensive research project.

\subsection{Centrality}

Using degree, eigenvector centrality, betweenness, closeness, and TOM, we rank the central importance of each of the individuals in the dataset.

\begin{figure}[ht]
\begin{center}
\includegraphics[scale=.5]{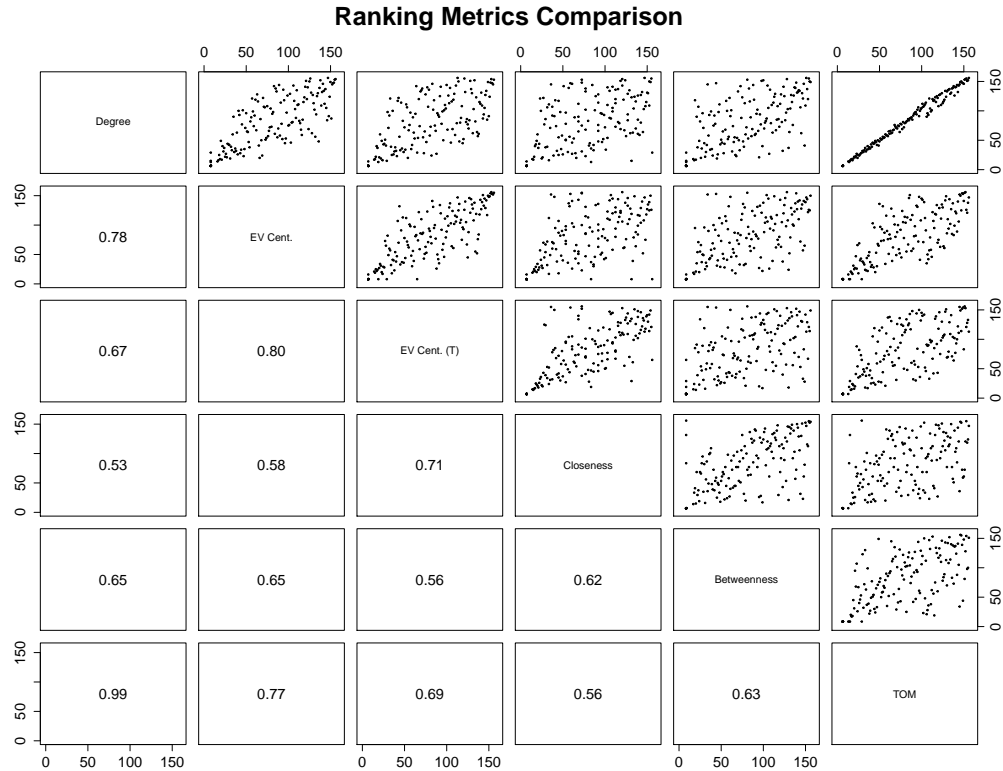}\\
\caption{For each of the measures of centrality, we find the ranked list of employees.  The ranked lists are then plotted against each other.  The number in the lower triangle represents the Pearson correlation associated with the comparison of the two relevant ranked lists.}\label{fig:pairs}.
\end{center}
\end{figure}

A student can spend considerable effort thinking about the different metrics used to rank the individuals in the network.  Recall, the more different people $i$ emails, directly or by cc, the greater $i$'s degree.  For instance, an employee who forwards a single announcement to everybody in the company can achieve maximal degree.  See figure \ref{fig:pairs} for a comparison of the centrality measures evaluated in this project.


\normalsize

\subsection{Network}

Using the R package Weighted Gene Co-expression Network Analysis (WGCNA) \citep{Lan08}, we cluster the observations into a hierarchical dendrogram. WGCNA uses a hierarchical clustering algorithm in an agglomerative (building one step at a time from 156 groups until all individuals are in one group) process to link individuals sequentially based on the number of emails exchanged.  We used average-linkage to determine closeness to a group that has already been formed; that is, an individual will be added to a group if they are close, on average, to the members of the existing group.  Additionally, we did not require that every individual be linked into a group.  We require that the dissimilarity be no more than 0.9 for the adjacency matrix. (Recall that the adjacency score is determined by the number of emails sent and received, divided by the maximum adjacency score.  The dissimilarity is one minus the adjacency.)  We require the TOM dissimilarity to be no more than 0.95. Lastly, each group is required to have at least 4 members according to our analysis.  The dissimilarity measure, linkage decision, and cutoff criteria are all parameters that can be adjusted in order to gain further insight into the data.

\section{Further Directions}\label{fd}

We presented above some suggested directions that students can take with discussion and research questions for each of the centrality measures.  We add to them here with some suggestions for class-specific modules and further exploration.

\subsection{Connections to Specific Courses}

\subsubsection{Introductory Statistics}
The analyses done in this article are not typically covered in Introductory Statistics.  However, the data could be used to do descriptive statistics.  For example, students could make boxplots across different Enron departments using either number of emails sent or number of emails received.  One might be able to run an inferential (e.g., chi-square) test to see if lawyers sent more emails to other lawyers or to non-lawyers.  Indeed, an interesting classroom discussion could be based on the data clearly {\sl not} being a representative sample from a population; instead, the data might be thought of as a sample from a process of email sending by the 156 individuals measured.

\subsubsection{Applied Statistics}
The data and analyses provided seem most appropriate for an applied statistics course (e.g., computational statistics, multivariate analysis, or data science) with an introductory prerequisite.  The Enron data allow for a complete analysis of centrality metrics as well as a consideration of different network or clustering construction methods which are based on distances.  We have provided R code for an initial analysis, but our work could easily be expanded to include additional centrality measures or other network and clustering construction methods.

\subsubsection{Mathematical Multivariate Analysis or Linear Algebra}  Principal component analysis is a mainstay of multivariate analysis classes, and increasingly, eigenvector centrality makes a late-semester appearance in linear algebra classes.  We submit that eigenvector centrality is at least equally as appropriate for a course in multivariate analysis in addition to, or instead of, PCA.  Both PCA and eigenvector centrality require some linear-algebraic sophistication and dexterity with eigentheory.  However, eigenvector centrality can be more intuitive---as the importance formula $x_i=\frac{1}{\lambda}\sum_{ij}x_j$ is a straightforward linear transcription of the importance-voting assumption of the model---while it still includes sophisticated machinery like the Perron-Frobenius Theorem.  On the other hand, the connection between eigenvectors of the covariance matrix and the principal axes of a best-fit ellipse can be obscure to the student upon first introduction.

\subsection{Alternative Applications \label{subsecAA}}
Some of the applications we suggest below might require direct manipulation of the email data, either to organize it differently or to compute different adjacency matrices. They might also require a database of employee titles and departments; \url{http://foreverdata.org/1009/Enron_Employee_Status.xls}.

\subsubsection{Data Cleanup}
To highlight the importance of data cleanup decisions, even if that is tangential to the focus of this paper or a course, students can discuss the multitude of ways to represent the Enron email network, and the potential consequences to the analysis of each decision or assumption made along the way.   For instance, are there employee-specific parameters that can be computed without constructing the entire network? Also, students can discuss different weightings for the matrix $M.$  What if being emailed directly and being cc-ed counted equally?  Is there a way to incorporate the importance of a message in the weighting, say by a blunt measure like the length of the email, or by a more sophisticated textual analysis?  Students would need to obtain all 500,000 emails with the information on {\sl From}, {\sl To}, and {\sl CC} \ fields of each email message; see section \ref{subsection:data} for additional details.

\subsubsection{Correlation between centrality and company hierarchy}  The managerial hierarchy of Enron is not reflected in the top ten employees as ranked by the centrality measures above.  Indeed, of the main executives at the company, only two appear in the top ten: Kenneth Lay, the CEO and chairman, came in fourth on the betweenness scale, and Greg Whalley, the president, had the eighth highest closeness score. While some studies have attempted to reconstruct the company hierarchy from the email network---see for instance \citet{GoldStandard} for an attempted recovery of dominance relationship from among the employees with known dominance-subordinate hierarchy by simply using the degree centrality---we are not aware of any studies that carefully interpret the significance of high rank in centrality measures in the context of the company's hierarchy.  Students would need at least the title information from each employee, see \citet{GoldStandard} for additional information.

\subsubsection{Gender and department} One of the interesting outcomes of our rankings is that the top eight scorers in eigenvector centrality were women.  Also, most of the top ten eigenscorers were lawyers.  There exists published studies that discuss email changes over time by department (see for instance \citet{Diesner05}), though they do not correlate the departments to the employees' centralities.  And while the Enron corpus has been used to study gender-related questions (like predicting gender from the email stream in \citet{Deitrick12}), we are not aware of centrality analyses of the Enron corpus with gender as a variable.  No additional data are needed for this extension.

\subsubsection{Generalized TOM and other centrality measures applied to TOM} As mentioned above, TOM can be generalized to $m$-step neighborhoods to measure agreement between nodes with respect to multiple steps of adjacency \citep{Yip07}.  Generalized TOM defines paths of length $m$ to define adjacency between nodes.  Additionally, a straightforward extension of TOM is to use other measures of adjacency (e.g., the binary measure of emails sent between two nodes) within the TOM metric.  Alternatively, applying centrality measure like eigencentrality or closeness to the TOM matrix instead of the graph adjacency matrix may result in deeper centrality measures that better take into account overall network connectedness.  No additional data are needed for this extension.

\subsubsection{Degree and Strength} A natural companion to degree centrality is strength. The {\sl strength} $\sigma_i$ of employee $i$ is defined to equal the total number of emails that $i$ sent or received.  For instance, we could compute $\sigma_i=\sum_j u_{ij}$. Like degree, strength is also a size measure, but of the volume of $i$'s correspondence instead of the extent of $i$'s network.  The more emails $i$ sends, the greater $i$'s strength. The degree $\delta_i$ and strength $\sigma_i$ of an employee $i$ are blunt centrality measures, but they can be effectively combined with a tuning parameter $\alpha$ to define the new centrality measure $\kappa_i(\alpha)=\delta_i^\alpha \sigma_i^{1-\alpha}$.  At an exploratory level, a student can vary $\alpha$ to observe corresponding differences in rankings.  A more sophisticated exploration might begin with asking whether there are critical $\alpha$ values that change the nature of the ranking in some fundamental way.  For instance, $\alpha=0$ corresponds to strength and $\alpha=1$ to degree.  Also, the range $0<\alpha<1$ seems to be fundamentally different from the range $\alpha>1$.  But are there less obvious critical values?  See \citet{Opsahl} for background on the tuning parameter.  No additional data are needed for this extension.

\subsubsection{Weights and directions}  All of our analysis was conducted on the weighted network under the assumption that a higher volume of emails must have more significance than a lower one.  But a simple unweighted graph of email connections, perhaps constructed with some minimum threshold for the number of emails, may reveal information that was obscured by the weighting.  Alternatively, students may gain insight from a kind of weighting that treats cc-ed employees differently from our reciprocal square root approach or that assigns importance to emails based on word count or sentiment analysis.   And additionally, whether the graph is directed or undirected---that is, whether the sender and receiver are treated symmetrically or not---will result in different outcomes for all the centrality measures, and each may suggest results that the other does not.  Students would need to obtain all 500,000 emails with the information on {\sl From}, {\sl To}, and {\sl CC} \ fields of each email message; see section \ref{subsection:data} for additional details.

\subsubsection{A time factor} The majority of the Enron corpus consists of emails from 1998 to 2002.  Our graph and corresponding matrices aggregate all the emails into one network.  However, it may make sense to consider how the email network changes over time, by month or by quarter.  For instance, can an anomaly detection on the network over time point out any changes that arose from scandal-related communication?  See \citet{TimeSeries} for some work in that direction.  Students would need to obtain all 500,000 emails with the information on {\sl From}, {\sl To}, and {\sl CC} \ fields of each email message; see section \ref{subsection:data} for additional details.

\subsubsection{Clustering Extensions}
Hierarchical clustering is only one network algorithm that uses adjacencies or distances to break up observations into groups.  Partitioning methods typically break the nodes up into groups that partition the units.  That is, each node will go into exactly one group. Partitioning Around Medoids (PAM) \citep{Kau90} iteratively allocates points to the group with the closest medoid (a measure of center based on the nodes themselves), recomputes the medoid, reallocates points, and repeats until no points need further swapping.  Partitioning methods have the disadvantage that the user is required to specify the number of clusters; however, silhouette width can be used to choose the optimal number of clusters \citep{Rous87}.

Another possible project for students is to use permutation methods to evaluate the significance of the resulting clustering output.  That is, one could create a null distribution of dendrograms resulting from permuted data. A senior project or research experience might have the students engage with different ways of measuring the distance from a null dendrogram to the observed dendrogram.  No additional data are needed for this extension.

\subsubsection{Visualizations}
Our research students were particularly interested in different visualizations of the data.  They used D3 graphics to create a dependency wheel and an interactive network image (see \url{http://enron-network.herokuapp.com/TOM}) \citep{Kay14}.  Using applications like Shiny (\url{http://shiny.rstudio.com/}) allows students to think about how best to communicate results, and the Enron data provides myriad opportunities for creative visualizations. No additional data are needed for this extension.

\subsubsection{Text Mining}
As a much larger extension, with the entire email corpus, a student project could involve text mining of the content of the emails or of the email subject lines.  There could also be a connection between some of the network results and a sentiment analysis of the words used within the emails themselves.

\subsection{Resources}  We have found the following websites useful for further exploration of the data as well as for processed and simplified datasets.

\begin{itemize}
\item \url{https://snap.stanford.edu/data/email-Enron.html}  Stanford Network Analysis Project network analysis and data mining library.

\item \url{http://bailando.sims.berkeley.edu/enron_email.html} UC Berkeley Enron Email Analysis Project, includes natural language processing annotation, visualization and clustering tool, and database representation for efficient querying.

\item \url{http://homes.cs.washington.edu/~jheer//projects/enron/v1/} Updated version of visualization and clustering tool by Jeff Heer from Berkeley website above.

\item \url{http://research.cs.queensu.ca/home/skill/otherforms.html} Processed forms of Enron data including word frequencies and time stamps

\item \url{http://cis.jhu.edu/~parky/Enron/} Another set of processed databases into simplified forms like (time, from, to) tuples.

\end{itemize}

\section*{Acknowledgements}

We are grateful to Theo Vassilakis and Jim Addler at Metanautix for their help and suggestions in getting this project started, the Pomona College Math Department for its continued support of undergraduate research, and the students of the Pomona College Undergraduate Research Circle during the Spring semester of 2014, Timothy Kaye, David Khatami, Daniel Metz, and Emily Proulx, for pushing the project to fruition.

\section*{Appendix}
As an appendix to this work we provide the dataset given in equation (\ref{data}).  We also provide a list of the 156 employees considered in the analysis (with their departmental affiliation and title).  The analysis was done using R (\url{http://www.r-project.org/}) and RStudio (\url{http://www.rstudio.com/}), and the code used for the analysis is provided as a markdown file and a pdf file.

\begin{itemize}
\item
The 156 x 156 adjacency matrix is available as a comma-separated value file: \url{http://www.amstat.org/publications/jse/.../Final Adjacency Matrix.csv}
\item
The list of 156 employees with their department affiliation and title is available as a comma-separated value file: \url{http://www.amstat.org/publications/jse/.../Enron Employee Information.csv}
\item
The R Markdown file including the code for the entire analysis is available at: \url{http://www.amstat.org/publications/jse/.../enronTutorial.Rmd}
\item
The associated pdf file compiled from the markdown code is available at: \url{http://www.amstat.org/publications/jse/.../enronTutorial.pdf}
\end{itemize}

\bibliographystyle{asa}
\bibliography{enron_JSE}

\vspace{-0.5in}

\section*{}
J.~S. Hardin\label{jo}\\
Pomona College\\
Department of Mathematics\\
610 North College Ave\\
Claremont, CA, 91711\\
\href{mailto:jo.hardin@pomona.edu}{\color{blue}\underline{Jo.Hardin@Pomona.edu}}

\vspace{-0.5in}

\section*{}
G. Sarkis\label{gs}\\
Pomona College\\
Department of Mathematics\\
610 North College Ave\\
Claremont, CA, 91711\\
\href{mailto:ghassan.sarkis@pomona.edu}{\color{blue}\underline{Ghassan.Sarkis@Pomona.edu}}

\vspace{-0.5in}

\section*{}
P.~C. URC\label{pc}{\protect\footnote{P.C.\ URC stands for the Pomona College Undergraduate Research Circle, whose members for this project were Timothy Kaye, David Khatami, Daniel Metz, and Emily Proulx.}}\\
Pomona College\\
Department of Mathematics\\
610 North College Ave\\
Claremont, CA, 91711\\
\href{mailto:PCURC@sakai.claremont.edu}{\color{blue}\underline{PCURC@Sakai.Claremont.edu}}

\end{document}